\documentclass[aps,prl,twocolumn,showpacs,floats,amsmath]{revtex4}%
\usepackage{graphicx}
\usepackage{amsmath}
\usepackage{amsfonts}
\usepackage{amssymb}%
\usepackage{epsf}

\newcommand{\be}{\begin{equation}}
\newcommand{\ee}{  \end{equation}}
\newcommand{\ba}{\begin{eqnarray}}
\newcommand{\ea}{  \end{eqnarray}}
\newcommand{\bs}{\begin{subequations}}
\newcommand{\es}{\end{subequations}}

\begin{document}

\title{Small polarons in dilute gas Bose-Einstein condensates}
\author{F. M. Cucchietti}
\author{E. Timmermans}
\affiliation{T-4, Theory Division, Los Alamos National
Laboratory, Los Alamos, NM 87545}
\date{\today}

\begin{abstract}
A neutral impurity atom immersed in a dilute 
Bose-Einstein condensate (BEC) can have a bound 
ground state in which the impurity is self-localized.  In this small polaron-like state, the 
impurity distorts the density of the surrounding BEC, thereby creating the
self-trapping potential minimum.  We describe the self-localization in
a strong coupling approach.

\end{abstract}

\pacs{03.75.Hh,67.40.Yv}

\maketitle

Experimentalists are pursuing the localization and transportation of individual 
atoms in dilute gas Bose-Einstein condensates (BECs) \cite{tweezer,Cas}.   
Their motives are manifold: 
The transportation of particles into and out of a localized state 
would realize a quantum-dot-like single particle device.  
The rate at which the localized state receives or emits particles could determine
the local density of states like a scanning tunneling microscope (STM).  
The motion of a localized atom could test superfluid dynamics \cite{Pitaevskii},
and its acceleration the Unruh effect \cite{unruh}.  
Light resonant with multiple localized particles could itself exhibit localization \cite{light}.  
The spins of localized particles could make up a quantum register of movable qu-bits.
However, the challenge of localizing a neutral atom by means of a steep
external potential is daunting \cite{Cas}.  In this Letter, we propose an
alternative strategy: an impurity self-localizes within a region smaller than the BEC-healing length
when the magnitude of the impurity-boson scattering length is increased above a critical value.
Similar to the electron self-localization in a polar crystal (forming a small polaron),
the BEC-impurity localizes because the interaction energy gain 
(stemming from the local distortion of the boson-field) outweighs the kinetic 
energy cost.  Observing this phenomenon in cold atoms may require a Feshbach 
resonance (to alter the impurity-boson interaction) and impurity creation 
(either by a Raman process or by trapping a different atom species),
but these techniques have been demonstrated \cite{impurities}.  This experiment
would create a novel class of self-localized particles: polarons with 
mass comparable to, or possibly larger than that of the boson particles.

In the context of condensed $^{4}$He-fluids, Miller et al.  \cite{Miller}
remarked on impurity self-localization and mentioned the polaron connection.
They advocated a perturbation treatment (weak-coupling theory)
by demonstrating the similarity of the perturbed wavefunction with
the variational one proposed by Feynman and Cohen to include
`backflow' \cite{Cohen}.  The weak-coupling description predicts that phonon-drag
increases the impurity mass from $m_{I}$ to $m_{I}^{\ast}=m_{I}/[1-\alpha]$,
where $\alpha$ is a dimensionless coupling constant.  Its value 
depends on the BEC density
$\rho_{B}$, the impurity-boson interaction potential $V_{q}$, the
boson mass $m_{B}$, and the excitation energy $E^{B}_{q}$ of the boson
modes of momentum $q$:
\be
\alpha = \frac{4}{3} \left( \frac{m_{B}}{m_{I}} \right) \frac{\rho_B}{(2\pi)^{3}}
\int d^{3} q \frac{\left|V_{q} (\hbar^2 q^{2}/2 m_{B})\right|^{2}}
{E^{B}_{q}[E^{B}_{q}+\hbar^{2} q^{2}/2m_{I}]^3} \; .
\ee
For a dilute BEC and an impurity-boson contact 
interaction of scattering length $a_{IB}$,
$V_{q} \rightarrow \lambda_{IB} = 2 \pi \hbar^{2} (1/m_{B}+1/m_{I}) a_{IB}$
and 
$E_{q}^{B} \rightarrow \hbar c_{B} q \sqrt{1 + (\xi q)^{2}}$.
Here, $\xi$ is the BEC-healing length which depends on the
boson-boson scattering length $a_{BB}$, $\xi=1/\sqrt{16 \pi \rho_{B} a_{BB}}$,
and $c_{B}$ denotes the BEC velocity of sound, $c_{B}=\hbar/(2 m_{B} \xi)$, so that
\be
\alpha = \frac{8}{3\sqrt{\pi}}
\sqrt{\frac{\rho_{B} a_{IB}^{4}}{a_{BB}}} \left( 1+\frac{m_{B}}{m_{I}} \right)^{2} 
\left( \frac{m_{B}}{m_{I}} \right)
F\left( \frac{m_{B}}{m_{I}} \right) \; ,
\ee
where 
$F(y) = \int_{0}^{\infty} x^{5}/\left[ \sqrt{1+x^{2}}(x \sqrt{1+x^{2}} + y x^{2})^{3} \right]$.
As in polaron physics, the effective mass divergence at 
$\alpha=1$ indicates self-localization, even though the weak-coupling description breaks down when $\alpha \sim 1$ \cite{Mahan}.

We describe the self-localized impurity in a strong-coupling treatment  
-- similar to the Landau-Pekar description of small polarons \cite{LandauPekar} --
using a product wavefunction:
\be
\Psi_{1,N}({\bf r},{\bf x}_1,{\bf x}_2,...,{\bf x}_N) \simeq \chi({\bf r})\psi({\bf x}_1)\psi({\bf x}_2)...\psi({\bf x}_N),
\label{product}
\ee
\noindent
where $\chi({\bf r})$ represents the impurity wavefunction and $\psi$ denotes the single particle 
state occupied by the $N$ indistinguishable bosons of position ${\bf x}_{j}$.  
We substitute Eq.(\ref{product}) into the Hamiltonian eigenvalue
equation and multiply first by $\psi^{\ast}({\bf x}_1)\psi^{\ast}({\bf x}_2)...\psi^{\ast}({\bf x}_N)$,
then by $\chi^{\ast}({\bf r})\psi^{\ast}({\bf x}_2)...\psi^{\ast}({\bf x}_N)$. Integrating the first
equation over  ${\bf x}_1 {\bf x}_2 ...  {\bf x}_N$, the second over
${\bf r} \; {\bf x}_2 ... {\bf x}_N$, and choosing the ground state wavefunction to be
real valued (e.g. $|\chi|^{2} = \chi^{2}$) we obtain \cite{Ryan}
\bs
\be
E_I \chi({\bf r}) = -\frac{\hbar^2 \nabla_{{\bf r}}^2}{2 m_I} \chi({\bf r}) + \lambda_{IB} \varphi^2({\bf r}) \chi({\bf r})
\label{FullEqImpurity}
\ee
\be
\mu_B \varphi({\bf x}) =  -\frac{\hbar^2 \nabla_{{\bf x}}^2}{2 m_B} \varphi({\bf x}) + \lambda_{BB} \varphi^3({\bf x}) + \lambda_{IB} \chi^2({\bf x}) \varphi({\bf x}),
\label{FullEqCondensate}
\ee
\label{FullEq}
\es
\noindent
where $\lambda_{BB}=[4\pi \hbar^{2}/m_{B}] a_{BB}$ and 
$\varphi$ is the condensate field, $\varphi=\sqrt{N}\psi$.
If $E_{1,N}$ and $E_{0,N}$ are the ground state energies 
of N bosons in the presence of one or zero impurity atoms, the
BEC chemical potential is $\mu_B =
E_{1,N}-E_{1,N-1}$, and the impurity energy is $E_I=E_{1,N}-E_{0,N}$.

Since the BEC experiences the density of only a single impurity, its field may be
altered only slightly for sufficiently weak boson-impurity coupling, $\varphi({\bf r}) = 
\sqrt{\rho_B^{0}} + \delta \varphi({\bf r})$.  Using $\mu_B\approx\lambda_{BB} \rho_B^0$, 
the corresponding linearization of Eqs.~(\ref{FullEq}) gives
\bs
\be
E_b \chi({\bf r}) = -\frac{\hbar^2 }{2 m_I} \nabla_{{\bf r}}^2 \chi({\bf r}) + 2 \lambda_{IB} \sqrt{\rho^0_B} \delta \varphi({\bf r}) \chi({\bf r})
\label{LinearEqImpurity}
\ee
\be
\left[ \nabla_{{\bf x}}^2 -\xi^{-2} \right] \delta \varphi({\bf x}) = \frac{2 m_B \lambda_{IB} \sqrt{\rho_B^0}}{\hbar^2} \chi^2({\bf x}),
\label{LinearEqCondensate}
\ee
\label{LinearEq}
\es
\noindent
where
$E_b=E_I-\lambda_{IB} \rho^0_B$ is the binding energy.

As a modified Helmoltz equation, we solve Eq. (\ref{LinearEqCondensate}) in terms of the
Green function $G_\xi({\bf r}) =(4 \pi)^{-1} e^{-|{\bf r}|/\xi} / |{\bf r}|$,
\be
\delta\varphi({\bf x}) = - \frac{1}{2\sqrt{\rho_B^0}} \frac{\lambda_{IB}}{\lambda_{BB} \xi^2} \int {\rm d}{\bf r} \ G_\xi({\bf x}-{\bf r}) \chi^2({\bf r}).
\label{deltaphi}
\ee
The excess number of BEC-atoms in the impurity region is then
$\int d{\bf r} \ 2 \sqrt{\rho_{B}^{0}} \delta\varphi({\bf r}) = - (\lambda_{IB}/\lambda_{BB})$, in
agreement with \cite{pethick2} where this number, induced by a general potential,
was determined from thermodynamic arguments.
The substitution of (\ref{deltaphi}) into 
(\ref{LinearEqImpurity}) results in the wave equation of a particle that self-interacts through
an attractive Yukawa (or screened Coulomb) potential.  Exploiting the Coulomb
analogy, we introduce an effective charge $Q$ where $Q^{2} = [\lambda_{I,B} \rho_{B}^{0}] a_{I,B}
\times 2 (1+m_B/m_I)$,
\begin{eqnarray}
E_{b} \chi({\bf r}) &=& -\frac{\hbar^2 }{2 m_I} \nabla_{{\bf r}}^2 \chi({\bf r})
\label{FinalEqImpurity} \\
&& - \left[  \int {\rm d}{\bf x} \frac{Q^{2}}
{|{\bf r}-{\bf x}|} e^{-\frac{|{\bf r}-{\bf x}|}{\xi}}  \; \chi^2({\bf x}) \right]  \chi({\bf r}).
\nonumber
\end{eqnarray}
Incidentally, $V_{\rm{med}}({\bf r}) = - Q^{2} e^{-|{\bf r}|/\xi }/|{\bf r}|$
is also the BEC-mediated  interaction experienced by distinguishable particles
immersed in a BEC, as  calculated from perturbation theory \cite{pethick,Stoof}.
The Coulomb analogy also suggests natural units $E_{0}$, the 
Rydberg energy, and $R_{0}$, an effective Bohr radius, $Q^{2}/R_{0}=
\hbar^{2}/m_{I}R_{0}^{2}= 2 E_{0}$,
\begin{eqnarray}
E_{0} &=&  \left[ \lambda_{IB} \rho_{B}^{0} \right] [a_{IB}^{3} \rho_B^{0}] \times 
4 \pi \left( \frac{m_{B}}{m_{I}} \right)^{2} (1+\frac{m_{I}}{m_{B}})^{3}
\nonumber \\
R_{0} &=& \frac{1}{4 \pi a_{IB}^{2} \rho_{B}^{0}} \times \frac{m_{I}m_{B}}{(m_{B}+m_{I})^{2}} \; ,
\label{natural}
\end{eqnarray}
which set the relevant energy ($E_{0}$), time ($\hbar/E_{0}$) and length ($R_{0}$) 
scales.  Note that $[4 \pi a_{IB}^{2} \rho_{B}^{0}]^{-1}$ is the mean free path of
an impurity moving among hard-sphere scatterers distributed at the BEC-density.

The small polaron then corresponds to solutions of (\ref{FinalEqImpurity}) with
negative eigenvalue $E_{b}$.  We break translational symmetry by hand and solve 
(\ref{FinalEqImpurity}) iteratively for an s-wave impurity wavefunction centered on the origin.  
At each iteration step, we solve the eigenvalue problem for an impurity particle experiencing
an effective potential  $u(r) =  \left[
\int {\rm d}{\bf x} \frac{-Q^{2}} {|{\bf r}-{\bf x}|} e^{-\frac{|{\bf r}-{\bf x}|}{\xi}} \;  \chi^2({\bf x})
\right]$, in which we substitute the impurity density $\chi^{2}({\bf x})$ obtained from
the previous iteration.  Spherical symmetry $u(r)$ to a one-dimensional integral.  
Working in natural units (\ref{natural})
and defining $\beta=\xi/R_{0}$, $u(r)$ reads
\ba
u(r)&=& \frac{-8 \pi \beta}{r}
\left[ e^{-r/\beta} \int_0^r {\rm d}r' \ {\rm sinh} \left( \frac{r'}{\beta} \right) \ r' \ \chi^2(r') \right. \nonumber \\
&+& \left. {\rm sinh} \left(\frac{r}{\beta}\right) \int^\infty_r {\rm d}r' \  e^{-r'/\beta} \ r' \ \chi^2(r') \right],
\label{OneDPotential}
\ea
where $\beta$ represents the only density/interaction dependence that
remains, thereby becoming the relevant dimensionless coupling constant, 
\be
\beta=\frac{\xi}{R_{0}}= \sqrt{\pi} \sqrt{\frac{a_{IB}^{4}}{a_{BB}} \rho_{B}^{0}} 
\left( 1 + \frac{m_{B}}{m_{I}} \right) \left( 1 + \frac{m_{I}}{m_{B}} \right) .
\ee
Another candidate, $Q^{2}/[\hbar c_{B}]$, is simply proportional, 
$Q^{2}/[\hbar c_{B}]= 2 (m_{B}/m_{I}) \beta$.  In Fig. 1 we show the iteratively obtained 
$\chi(r)$-functions, whereas the 
inset shows the corresponding binding energies (in units of $E_{0}$), for several $\beta$-values.  
Thus, this strong-coupling description predicts 
the impurity self-localizes if $\beta > 4.7$.
The deeply bound variational wave function ($\beta > 20)$ with width $\sigma = 3 \sqrt{\pi/2}$,
shown by the bold line of Fig. 1, is remarkably similar to the iterative function \cite{remark}.

\begin{figure}[tb]
\centering \leavevmode
\epsfxsize 3.2in
\epsfbox{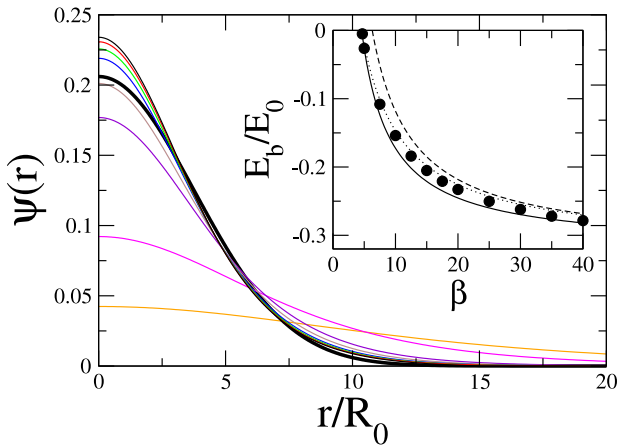}
\caption{Radial wavefunction obtained through the iterative procedure for
$\xi/R_0=4.7, 5, 10, 20, 30, 40$ (from bottom to top).
In bold black line, the initial Gaussian guess.
In the inset, the energy of the ground wavefunction vs $\xi$ (dots).
In dotted line, the variational result obtained numerically. The expansion
for large $\xi/R_0$ is in dashed line, and the best fit
$E_{b}/E_0 \simeq -1/\pi + 3 R_0/2\xi$ in solid line.}
\label{Figure1}
\end{figure}

When $\beta\geq20$, the impurity
state converges to that of a particle bound by a pure Coulomb-self
interaction with energy $E_{b} = - 0.316 E_{0}$ and extent $\sqrt{\langle r^{2} \rangle}
= 4.64 R_{0}$.

To understand the interesting `transition' regime, $4.7 < \beta \leq 20$, which exhibits 
an intricate interaction dependence, we approximate the impurity 
wavefunction variationally.  The effective impurity equation (\ref{FinalEqImpurity}) 
is equivalent to minimizing the functional $E_{V} = T + V/2$ \cite{variationalremark}, where $T$ denotes 
the kinetic energy
$T=-(\hbar^{2}/2m_{I}) \int {\rm d}{\bf r} \chi({\bf r}) \nabla^{2} \chi({\bf r})$ and
$V$ the self-interaction energy $V= \int {\rm d}{\bf r} \chi^{2}({\bf r}) u({\bf r})$
, with respect to variations of the
real-valued normalized wavefunction, $\chi({\bf r})$.  Choosing a Gaussian
trial wavefunction, 
$\chi({\bf r})~=~\exp{(-|{\bf r}|^2/2 \sigma^2)}/(\pi \sigma^2)^{3/4}$,
the functional, written in natural units, becomes
\be
E_{V} = \frac{3}{2 \sigma^{2}} - \sqrt{\frac{2}{\pi}} \frac{f(\sigma/\beta)}{\sigma} \; ,
\label{variational}
\ee
where $f(a) = \int dr \ r e^{-ra} e^{-r^{2}/2}$. Numerical minimization of (\ref{variational})
with respect to $\sigma$ gives a binding energy that agrees well with the iterative 
solution of (\ref{FinalEqImpurity}), shown in dotted line in the inset of Fig. 1. The dashed 
line plots the energy obtained by expanding (\ref{variational}) for large $\beta$, 
$E_{b}/E_0 \simeq -1/\pi + 2/\beta$, showing reasonable agreement with the iterative 
solution but slightly overestimating the minimal $\beta$-value for
self-localization.  A fit that gives better agreement over the whole $\beta$-range is 
$E_{b}/E_0 \simeq -1/\pi + 1.5/\beta$  (solid line in inset of Fig. 1).  
With $E_{0} = [\lambda_{BB} \rho_{B}^{0}] 2 (m_{B}/m_{I}) \beta^{2}$, 
$E_b$ also equals
\be
E_{b} \approx [\lambda_{BB} \rho_{B}^{0}] 2 (m_{B}/m_{I})
\left[ \frac{3}{2} \beta - \frac{\beta^{2}}{\pi} \right],
\label{bind}
\ee
reminiscent of the strong coupling energy of traditional polarons, proportional to the square
of the coupling constant \cite{Mahan}.

\begin{figure}[t]
\centering \leavevmode
\epsfxsize 3.2in
\epsfbox{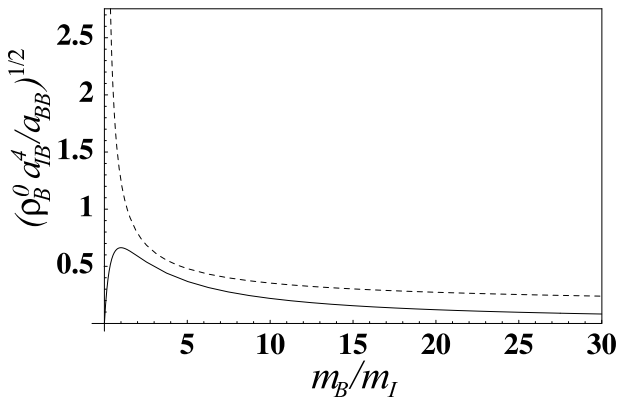}
\caption{Minimal $\sqrt{\rho_{B}^{0} a_{IB}^{4}/a_{BB}}$-value
for localization of the impurity as a function of the mass ratio $m_{B}/m_{I}$ obtained
from weak coupling ($\alpha>1$) \cite{Miller} and strong coupling 
(Eq. (\ref{FinalEqImpurity}) and $\beta>4.7$) descriptions, dashed and solid line respectively. }
\label{Figure2}
\end{figure}

In Fig. 2 we compare the corresponding minimal $\sqrt{\rho_{B}^{0} a_{IB}^{4}/a_{BB}}$-value
for localization as a function of $(m_{B}/m_{I})$ predicted by the weak $(\alpha>1)$ \cite{Miller}
and strong $(\beta >4.7)$ coupling descriptions.  Although neither treatment should be
quantitatively correct, they give comparable results for $1<(m_{B}/m_{I})<10$.

\begin{figure}[b]
\centering \leavevmode
\epsfxsize 3.2in
\epsfbox{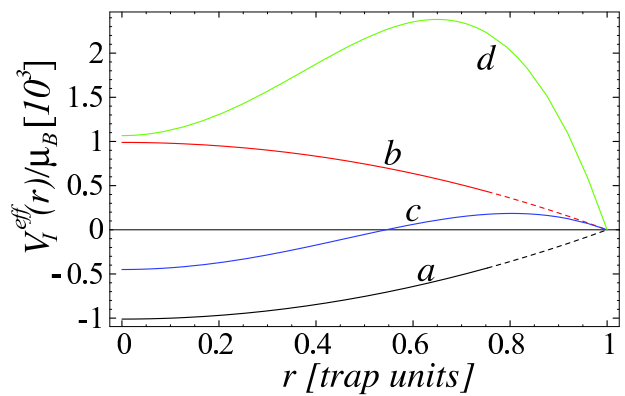}
\caption{Effective potential of the impurity as a function of distance to the trap center for
(a) attractive and (b) repulsive boson-impurity interaction, using $\rho_B^0(0)|a_{IB}^3|=10^{-3}$,
$|a_{IB}|/a_{BB}=10^3$ and $m_I=m_B$. (c) and (d) have $a_{IB}>0$, but the former has
$\rho_B^0(0)|a_{IB}^3|=0.05$ and $m_I=m_B$, while in the latter $\rho_B^0(0)|a_{IB}^3|=10^{-3}$ and $m_B=14.5 m_I$($^6 Li$ impurities in a $^{87}Rb$ BEC).
The dashed lines show the values at which the localization condition, $\beta>4.7$, is not fulfilled.  }
\label{Figure3}
\end{figure}

In an inhomogeneous BEC confined by a trapping potential
$V_{B}({\bf r})$, the impurity-free density $\rho_{B}^{0}({\bf r})$ varies spatially.
Assuming that $\rho_{B}^{0}({\bf r})$ varies slowly on the
scale of $R_{0}$ ($R_{0} |\nabla \rho_{B}^{0}|/\rho_{B}^{0} <<1$), we describe the
self-localized impurities (which appear point-like to the BEC) as immersed
in a locally homogeneous superfluid.  If the impurities localize on a time scale
shorter than the time for the impurity to move appreciable ($E_{0}/\hbar>>\omega_{trap}$), 
or for the impurities to attract each other (which depends on the average impurity density), 
we can describe the subsequent impurity dynamics as that of classical point particles
subject to an effective potential.  This potential energy $V_{I}^{eff}({\bf r})$
is the sum of the external impurity potential $V_{I}^{ext}({\bf r})$, the mean-field 
energy $\lambda_{IB} \rho_{B}^{0}({\bf r})$, and the local binding energy 
$E_{b}[\rho_{B}^{0}({\bf r})]$ of Eq.(\ref{bind}).  Computing $\rho_{B}^{0}({\bf r})$ in the 
Thomas-Fermi approximation, we find
\begin{eqnarray}
&&V_{I}^{eff}({\bf r}) = V_{I}^{ext}({\bf r}) + \mu_{B} \left[ 1 - V_{B}({\bf r})/\mu_{B} \right]
\times
\nonumber \\ 
&& \left\{ \frac{\lambda_{IB}}{\lambda_{BB}} + 2 \frac{m_{B}}{m_{I}} 
\left( \frac{- \beta^{2}({\bf r})}{\pi} + \frac{ 3 \beta({\bf r})}{2}  \right) \right\} \; ,
\end{eqnarray}
where
\begin{eqnarray}
\beta ({\bf r}) &=& \sqrt{\pi} 
\left( 1 + \frac{m_{B}}{m_{I}} \right) \left( 1 + \frac{m_{I}}{m_{B}} \right)
 \nonumber \\ 
& & \ \sqrt{\frac{a_{IB}^{4}}{a_{BB}} \rho_{B}^{0}({\bf r}=0)}
 \times \sqrt{1 - \frac{V_{B}({\bf r})}{\mu_{B}}}.
\end{eqnarray}
Even when $V_{I}^{ext}=0$ and the impurity-BEC interaction is repulsive $\lambda_{IB}>0$
-- so that the boson mean-field (the first term in the $\left\{ \right\}$-bracket)
would expell the impurity from the trap center -- the binding energy (the other terms in the 
$\left\{ \right\}$-bracket) can give an overall potential that 
attracts the impurities to the trap middle.  This behavior is illustrated in Fig. 3 
for typical experimental parameters for $^6 Li$ impurities in a $^{87}Rb$ BEC
($m_B/m_I = 14.5$).  Even if in the true ground state the impurity would hover
at the edge of the BEC, the self-localization can form a metastable state with long
tunneling times.  In any case, $V_{I}^{ext}$ can keep the impurities within the
BEC, and $\lambda_{BI}>0$-impurities tend to gather in the trap center.

A question remains regarding the accuracy of the product state (\ref{product})
when $m_B/m_I \sim 1$, although $|E_b|\gg \mu_B$ implies a separation of
time scales that justifies the lack of impurity-BEC correlations.
For $|E_b|\lesssim \mu_B$, a more sophisticated description could be useful:
we  expect the above results to serve as a benchmark for future calculations.

When is the linearization of $\varphi$ justified? With $\chi({\bf r}) = e^{-r^{2}/2\sigma^{2}}/
(\pi \sigma^{2})^{3/4}$, the ratio of the peak-value of the impurity-induced fluctuation,
$\delta \varphi({\bf r}=0)$, to its spatial average, $\sqrt{\rho_{B}^{0}}$, equals
$\delta \varphi({\bf r}=0)/\sqrt{\rho_{B}^{0}} = -(1/\sqrt{\pi}) (1+ m_{B}/m_{I}) (a_{IB}/\sigma)
f(\sigma/[\sqrt{2} \xi])$.  Assuming a deeply bound polaron, $f(\sigma/[\sqrt{2} \xi]) \approx 1$
and $\sigma \approx 3 \sqrt{\pi/2} R_{0}$, the condition 
$|\delta \varphi({\bf r}=0)/\sqrt{\rho_{B}^{0}}|<1/10$ takes the form
\be
|\rho_{B}^{0} a_{IB}^{3}| < \frac{m_{I}^{2} m_{B}}{(m_{B}+m_{I})^{3}} \frac{1}{1.89} 
\frac{1}{10} \; .
\ee
A large increase in $a_{IB}$ above the critical value for self-localization
could collapse the system when the linearization condition 
$\delta \rho_{B}/\rho_{B}^{0}$ is violated, as found in \cite{Ryan}. We speculate
that in this regime $\lambda_{IB}>0$ impurities could `phase separate',
creating a hole in the BEC-density.

In addition, the self-localization condition, 
$\beta > 4.7$, gives a lower bound to $|\rho_{B}^{0} a_{IB}^{3}|$ and
\be
\frac{m_{I}^{2} m_{B}}{(m_{B}+m_{I})^{3}} \frac{1}{18.9} > |\rho_{B}^{0} a_{IB}^{3}| >
7.0 \left( \frac{a_{BB}}{a_{IB}} \right) \frac{m_{I}^{2} m_{B}^{2}}{(m_{B}+m_{I})^{4}} \; ,
\label{ineq1}
\ee
defines the regime in which the linearization assumption holds and 
the strong-coupling description predicts self-localization.
Furthermore, the $(a_{IB}/a_{BB})$ ratio
is subject to conditions stemming from the validity of the contact description of the
impurity-boson interactions, $|\rho_{B}^{0} a_{IB}^{3}|<<1$, and from the outer-most
inequalities of (\ref{ineq1}), giving, respectively
\be
\left| \frac{a_{IB}}{a_{BB}} \right| >> \frac{7.0 m_{I}^{2} m_{B}^{2}}{(m_{B}+m_{I})^{4}}, \ \ 
{\rm and} \ \left| \frac{a_{IB}}{a_{BB}} \right| > \frac{132 m_{B}}{(m_{B}+m_{I})}.
\ee
These conditions may require a Feshbach resonance, but this can 
be achieved with existing technology.  

	Time of flight measurements or diffraction of light resonant with 
impurities can detect small polaron formation.  In the former case, the tightly
bound impurity wave function can expand faster and further than if the impurity
were not self-bound \cite{Boshier}.  In the latter case, the opening angle
$\theta$ of the cone in which light of momentum $k$ is scattered coherently,
$2 \sin(\theta/2) \sim 1/[k \sqrt{\langle r^{2} \rangle}]$, for $k \sim R_{0}^{-1}$
widens abruptly when the impurity self-localizes.

	In summary, we have pointed out that a neutral impurity atom immersed in a 
homogeneous (or large) BEC can
self-localize in a region smaller than the BEC-healing length.  In a strong-coupling description
with BEC linearization, the localizing BEC-distortion gives rise to an attractive 
self-interaction with a spatial dependence identical to the BEC mediated impurity-impurity interaction.  
Roughly, binding occurs when the range of
the self-interaction range exceeds the extent of the bound impurity -- more
precisely, when $\beta > 4.7$, a condition that can be fulfilled experimentally.
Using a variational Gaussian impurity wavefunction, we construct an analytical
approximation to the binding energy from which we obtained the effective potential
energy experienced by self-localized impurities in a trapped BEC.


\begin{thebibliography}{99}
\bibitem{remark} The success of the Gaussian wavefunction is paralleled in the description
of traditional small polarons, see \cite{Mahan}.
\end{thebibliography}
\end{document}